\documentclass[11pt]{article}
\usepackage{acl}
\usepackage{times}
\usepackage{latexsym}
\usepackage[T1]{fontenc}
\usepackage[utf8]{inputenc}
\usepackage{microtype}
\usepackage{inconsolata}
\usepackage{graphicx}
\usepackage{booktabs}
\usepackage{amsmath}
\usepackage{newtxmath}
\usepackage{xcolor}

\title{Do Simulated Users Push Back Like Real Users? \\ Correctness vs.\ Stance}
\author{Xindi Huang \\ The University of Sydney}

\newcommand{\Dsim}{\Delta_{\text{sim}}}
\newcommand{\Dreal}{\Delta_{\text{real}}}

\hypersetup{pdftitle={Do Simulated Users Push Back Like Real Users? Correctness vs. Stance},pdfauthor={Xindi Huang}}
\begin{document}
\maketitle

\begin{abstract}
LLM-simulated users stand in for people in multi-turn evaluations, on the assumption that they react to what the assistant says. We test this for pushback. From WildChat, we take conversations in which the assistant explicitly challenges or endorses a position the user stated, and compare the real next user turn with simulated turns from the same prefix. Among users who replied, real users push back more after a challenge than after an endorsement (20.5\% vs.\ 8.2\%); a GPT-based simulator does not (27.1\% vs.\ 27.1\%): stance gap $D$\,=\,$-$0.12 [$-$0.17,~$-$0.08]. After a challenge the simulator instead often re-states or re-interprets its request (``I meant \ldots''); about half of these turns are disguised insistence, which puts $D$ between $-0.08$ and $-0.05$. What drives the simulator's disputing is whether the assistant's position is judged correct (21\% vs.\ 49\% after challenges; 16\% vs.\ 58\% after endorsements), while real users change little (19\% vs.\ 24\%; 7\% vs.\ 12\%), under every outcome definition we tried. Real same-stance examples do not fix this, and another GPT-based simulator and two Claude simulators show stance effects close to real users. Stance sensitivity should be measured for each simulator, not assumed.
\end{abstract}

\section{Introduction}
LLM user simulators stand in for people in multi-turn evaluations, including sycophancy benchmarks in which a simulated user challenges the model \citep{tang2026,pulipaka2026,parikh2026}. Such benchmarks are only as valid as the simulator's reactions. Simulators are known to be miscalibrated in \emph{level}: more cooperative and verbose, and less likely to leave, than real users \citep{ivey2026,zhou2026}. Level fidelity does not imply \emph{effect} fidelity \citep{li2026}: a simulator can match how often users push back without changing that rate when the assistant's behaviour changes.

We ask whether simulators reproduce real users' sensitivity to the assistant's \textbf{stance}: whether it \emph{challenges} or \emph{explicitly endorses} a position the user stated as their own. H1: real users push back more after a challenge. H2: simulators change their pushback less ($D<0$). Following \citet{ivey2026}, we compare the real next turn in WildChat \citep{zhao2024} with simulated next turns from the same prefix. Contributions:
(1) \textbf{the stance gap $D$}, which separates sensitivity from level on the same prefixes;
(2) \textbf{evidence from real conversations}: our primary GPT simulator changes its pushback far less than real users and re-states its request instead, while a second GPT simulator and two Claude simulators come closer to real users;
(3) \textbf{correctness more than stance} (exploratory): the GPT simulator's disputing follows whether the assistant's position is judged incorrect far more than whether the assistant disagrees, and it even corrects the position it is meant to hold;
(4) \textbf{two fixes}: real same-stance examples do not help; a knowledge-limiting prompt restores the stance effect without removing fact-checking.

\section{Related Work}
\paragraph{Simulator fidelity.} \citet{ivey2026} compare simulated and real third turns in WildChat on style; \citet{zhou2026} find simulators over-cooperative on $\tau$-bench; \citet{seshadri2026} show the simulator choice shifts agent success; \citet{chen2026} finds simulated customers compress real group differences, and \citet{liu2026synthetic} propose a framework comparing simulated and real users in multi-turn conversations; we test a conditional effect on the same real prefixes. \citet{li2026} argue for judging simulations by effects; \citet{kirk2026} find similar slopes for participants and their simulated twins. Real next turns signal (dis)satisfaction \citep{donyehiya2024,liu2025}.

\paragraph{Knowing too much and sycophancy.} Simulated participants answer more accurately than people \citep{aher2023}; role-played characters leak knowledge \citep{ahn2024}; simulated searchers and pupils show a ``curse of knowledge'' \citep{ma2026,balog2026}. Sycophancy work studies the \emph{model} after a user disagrees \citep{hong2025}; we study the \emph{user} after the model disagrees.

\section{Data and Method}
\paragraph{Data and probes.} We use WildChat-4.8M (ODC-BY; revision c827c6d), with conversations from April 2023 to July 2025; 20\% of conversations are held out for development. A probe is an assistant reply that takes a stance on a position the user stated as their own, outside an ongoing dispute. Challenge (C) probes come from the top 0.6\% of a TF-IDF ranker, endorse (E) probes from a second ranker and a GPT prescreen; Claude Sonnet~5.5 assigns final stance labels, blind to the next turn. C probes are thus clear challenges (in a random sample of 50 challenges, real pushback was 8--10\%, 95\% CI about 2--22\%). With one probe per conversation we have 868 C and 953 E (2023: 870; 2024: 718; 2025: 233); 1,777 (855 C, 922 E) with a real and a simulated reply form the main sample.

\paragraph{Simulators.} Given the prefix, a simulator writes the user's next message or \texttt{[NO\_REPLY]}, with a prompt adapted from \citet{ivey2026} (prompt released with our code). We use gpt-6-luna (Luna~6, or Luna; primary; all probes) and gpt-6.1-sol (Sol~6.1; all probes) through the Codex CLI, $K=3$ samples each (highly similar: median character 5-gram Jaccard 0.20), and, exploratorily, Claude Sonnet~5.5, Opus~5.5 and Haiku~4.5 on 200 probes.

\paragraph{Reactions.} An LLM judge (gpt-6.1-sol, temperature 0 requested; the relay may not honour it) labels each next turn alone, without its source. \emph{Pushback} means the user doubts, disputes or rejects what the assistant said, or insists on their position (codebook released with our code). The judge agrees with Claude Opus~5.5 at $\kappa=0.63$--$0.75$ across five sets of 100 conversations and with itself at $\kappa=0.91$, and 88\% of the turns it calls pushback carry human negative-feedback labels \citep{liu2025}; its recall on a public disagreement set \citep{cai2020} is about 0.45 (Appendix~A).

\paragraph{Stance gap.} For each source, $\Delta = P(\text{pushback}\mid C) - P(\text{pushback}\mid E)$, and $\mathbf{D = \Dsim - \Dreal}$; $D<0$ means the simulator changes less than real users, and a constant level bias cancels. The primary test is $D$ for Luna, replied-only, with 95\% bootstrap intervals over conversations (2,000 resamples), fixed in a written plan before the final judge run (after preliminary results with an earlier judge; timestamped commit in our code repository). Four secondary tests are Holm-adjusted: all-probes $D$ (no reply = no pushback; 300 random probes), $D$ for acknowledgement, user-clustered $D$, and method B vs.\ control. Everything else is exploratory.

\begin{figure}[t]
\centering
\fbox{\parbox{0.95\columnwidth}{\footnotesize
\textbf{User} (paraphrased): seen from outside, nothing ever falls into a black hole, because crossing the horizon takes forever while evaporation takes finite time.\\
\textbf{Assistant} (endorses): agrees that an outside observer never sees the object cross.\\[2pt]
\textbf{Real user:} restates the argument: the crossing takes infinite time, evaporation is shorter, so the hole always evaporates first. \textit{(keeps the claim)}\\
\textbf{Luna:} objects that the infinite time is an artefact of the coordinate system, not a real delay, so the argument fails. \textit{(corrects the user's own claim)}}}\\[4pt]
\includegraphics[width=0.95\columnwidth]{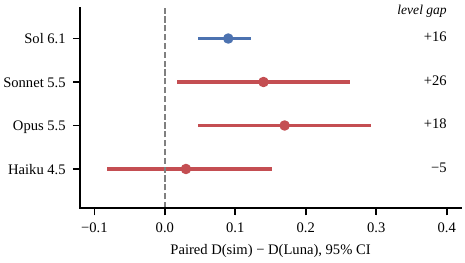}
\caption{Top: an endorsement of a claim judged incorrect (paraphrased). The real user keeps the claim; the simulator disputes the position it plays. Bottom: stance gap of each simulator relative to Luna on the same conversations (paired $D_{\text{sim}}-D_{\text{Luna}}$, 95\% CI; positive = more stance-sensitive than Luna), with the level gap (mean simulated minus real pushback, points). Red: Claude simulators (200 conversations); blue: Sol 6.1 (all conversations).}
\label{fig:1}
\end{figure}

\begin{table*}[t]
\centering\footnotesize\setlength{\tabcolsep}{4pt}
\begin{tabular}{lccccc}
\toprule
Simulator (n C/E) & Real C/E & Sim C/E & $D$ [95\% CI] & Paired vs Luna & Level gap \\
\midrule
Luna, all (855/922) & 20.5 / 8.2 & 27.1 / 27.1 & $-0.12$ [$-0.17$, $-0.08$] & -- & $+12.8$ \\
Luna, all-probes (150/150) & 14.0 / 6.7 & 23.8 / 33.3 & $-0.17$ [$-0.27$, $-0.08$] & -- & $+18.2$ \\
Luna, repair only & 5.7 / 2.8 & 21.2 / 3.5 & $+0.15$ [$+0.12$, $+0.18$] & -- & -- \\
Luna, pushback $\cup$ repair & 26.2 / 11.1 & 48.3 / 30.5 & $+0.03$ [$-0.02$, $+0.07$] & -- & -- \\
Luna, acknowledgement & 2.1 / 5.5 & 15.9 / 26.0 & $-0.07$ [$-0.10$, $-0.03$] & -- & -- \\
\midrule
Sol 6.1 (867/941) & 19.7 / 7.9 & 33.8 / 25.6 & $-0.04$ [$-0.08$, $+0.01$] & $+0.09$ [$+0.05$, $+0.12$] & $+15.9$ \\
Sonnet 5.5 (98/96) & 21.4 / 9.4 & 50.0 / 33.0 & $+0.05$ & $+0.14$ [$+0.02$, $+0.26$] & $+26.1$ \\
Opus 5.5 (99/96) & 21.2 / 9.4 & 43.4 / 23.6 & $+0.08$ & $+0.17$ [$+0.05$, $+0.29$] & $+18.2$ \\
Haiku 4.5 (99/96) & 21.2 / 9.4 & 13.5 / 7.1 & $-0.05$ & $+0.03$ [$-0.08$, $+0.15$] & $-5.0$ \\
Luna + examples, B (210/188) & -- & 24.0 / 23.5 & \multicolumn{2}{c}{B $-$ control: $-0.02$ [$-0.07$, $+0.04$]} & \\
\bottomrule
\end{tabular}
\caption{Pushback (\%) by stance and stance gap $D$, replied-only, single-turn judge. Level gap = mean of the C and E simulated rates minus the mean of the real rates (points). Paired = $D_{\text{sim}}-D_{\text{Luna}}$ on the same conversations. Rows from Sol 6.1 down are exploratory; the repair, pushback $\cup$ repair and acknowledgement rows use other outcome labels.}
\label{tab:1}
\end{table*}

\section{Results}
\paragraph{H1 and H2.} Among users who replied, real users push back after 20.5\% of challenges and 8.2\% of endorsements ($\Dreal=+0.12$); Luna after 27.1\% of both: $\mathbf{D=-0.12}$ [$-0.17$, $-0.08$], $p<0.001$ (Table~\ref{tab:1}). The secondary pushback tests agree (Holm $p=0.002$): counting users who left, $D=-0.17$ [$-0.27$, $-0.08$] (real users leave slightly more after a challenge, Luna after an endorsement); resampling users, $-0.12$. The acknowledgement test points the other way ($D=-0.07$ [$-0.10$, $-0.03$]): Luna acknowledges 10 points less after a challenge, real users 3 points. $D$ is similar in 2023 ($-0.14$) and 2024--25 ($-0.12$) and after regression adjustment ($-0.13$; Appendix Table~A1).

\paragraph{Pushback, not repair (exploratory).} After a challenge, 21.2\% of Luna's turns say the assistant misread the request (``I meant \ldots''), which our codebook labels repair (3.5\% after an endorsement; real users 5.7\% / 2.8\%). Reading 40 such turns (coded by an LLM auditor, Claude, not by a human), 19 re-interpret the request, often inventing a referent the user never mentioned, and 19 insist on it in other words, at least 12 of which the codebook would count as pushback. With a share $f$ of repairs that are really pushback, $D = 0.148f - 0.122$,\footnote{$0.148 = (0.212-0.035)-(0.057-0.028)$: the stance difference in repair rates for Luna minus that for real users.} i.e.\ $-0.08$ to $-0.05$ for $f=0.3$--$0.5$; counting all repairs gives $+0.03$ [$-0.02$, $+0.07$]. Simulators never change topic (real users 6--10\%), which our prompt likely induces.

\begin{table}[t]
\centering\footnotesize\setlength{\tabcolsep}{3pt}
\begin{tabular}{llrccc}
\toprule
Stance & Judged & n & Real & Luna & Opus lab.\ (R/L) \\
\midrule
C & correct & 649 & 19.3 & 21.1 & 13.8 / 25.4 \\
C & incorrect & 185 & 23.8 & 48.8 & 26.6 / 49.5 \\
E & correct & 637 & 7.2 & 16.2 & -- \\
E & incorrect & 214 & 12.2 & 58.1 & -- \\
\bottomrule
\end{tabular}
\caption{Pushback (\%) by stance and judged correctness of the assistant's position (exploratory, post hoc; replied-only). Correctness effect (incorrect $-$ correct): real $+0.05$ / $+0.05$, Luna $+0.28$ / $+0.42$ (C / E); Luna $-$ real $p\le0.001$ for both. Last column: challenges re-labelled by Claude Opus~5.5 (n = 80 / 64). ``Incorrect'' for a challenge often means the objection or refusal was unwarranted, not only factually wrong. CIs in Appendix.}
\label{tab:2}
\end{table}

\paragraph{$D$ separates simulators (exploratory).} On all conversations, Sol~6.1 gives $D=-0.04$ [$-0.08$, $+0.01$] and is more stance-sensitive than Luna on the same conversations (paired $+0.09$ [$+0.05$, $+0.12$]), though with even more excess pushback than Luna (level gap $+15.9$ vs.\ $+12.8$ points); like Luna, it follows judged correctness (26\% vs.\ 62\% pushback after challenges judged correct vs.\ incorrect; 18\% vs.\ 51\% after endorsements). On 200 conversations, Claude Sonnet~5.5 and Opus~5.5 are more stance-sensitive than Luna (paired $+0.14$ [$+0.02$, $+0.26$] and $+0.17$ [$+0.05$, $+0.29$]; $+0.10$ and $+0.11$, CIs including 0, when repair counts), with $D$ close to real users ($+0.05$, $+0.08$) but far more pushback (up to 50\% after challenges). Haiku~4.5 cannot be distinguished from Luna or real users. A Claude Opus~5.5 judge on 100 conversations per Claude simulator gives the same picture ($\kappa=0.63$--$0.74$ with ours). Sonnet and Opus also push back more when the assistant's position is judged incorrect: they follow both stance and correctness. Level and sensitivity are thus separate (Figure~\ref{fig:1}).

\paragraph{Correctness more than stance (exploratory, post hoc).} An LLM judge rated whether the assistant's position was correct: for challenges its disagreement, for endorsements the user position it endorsed. Real users' pushback follows stance whether the position is judged correct (19.3\% vs.\ 7.2\%) or incorrect (23.8\% vs.\ 12.2\%). Luna's pushback follows judged correctness: 21.1\% and 16.2\% when correct, 48.8\% and 58.1\% when incorrect (Table~\ref{tab:2}; differences from real users $p\le0.001$), also when repair counts as disagreement (Table~A1). Luna thus often disputes an endorsement of the very position it plays; in 20 such conversations read by the same LLM auditor, it corrected the user's own claim in 10 and no real user did (Figure~\ref{fig:1}). Because most positions are judged correct, Luna's overall $\Delta$ is near zero. Luna's pattern holds under Claude Opus~5.5 correctness labels on 150 challenges ($\kappa=0.52$; 25\% vs.\ 50\%), though there real users also differ (14\% vs.\ 27\%).

\section{Two Fixes: Examples Do Not Help, a Knowledge Bound Helps for the Wrong Reason}
\paragraph{Method B} shows, before the prompt, the real next turns of the three most similar same-stance probes from a disjoint pool (59 challenge and 163 endorsement examples used); a \textbf{control} shows three random real next turns. On 398 test probes, Luna's stance effect is $+0.00$ with same-stance and $+0.02$ with random examples (paired $-0.02$ [$-0.07$, $+0.04$], Holm $p=0.58$). Both bring the pushback level closer to real users (about 24\%) but not the sensitivity: examples teach how users talk, not when they push back.

\paragraph{A knowledge bound (exploratory).} We told Luna to use only what the user has shown and to react to whether the assistant agrees, on 360 paired conversations stratified by stance and judged correctness (criteria written before the run). The correctness effect fell from 0.41 to 0.31 ($-0.10$ [$-0.16$, $-0.04$]), short of the halving we set, and $D$ rose by $+0.11$ [$+0.04$, $+0.17$], almost all from endorsements of incorrect positions (pushback 56.6\%$\,\to\,$41.3\%; incorrect challenges 53.7\%$\,\to\,$52.0\%). On this sample the bound restores the stance effect ($\Dsim$ 0.12 vs. real 0.10), but not by removing fact-checking: it stops the simulator from correcting its own position, not from disputing a challenge it judges wrong. As the prompt also named the stance cue and \texttt{NO\_REPLY} rose from 3.5\% to 13.1\%, the gain may be instruction following; prompting alone does not bound knowledge \citep{balog2026}, and prompt interventions can shift the implicit simulated population \citep{lin2026}.

\section{Discussion and Conclusion}
Real users push back more when the assistant challenges them; our primary GPT simulator does not. It disputes what it judges wrong and answers a challenge by re-stating or re-interpreting its request. This is consistent with a ``curse of knowledge'' \citep{camerer1989}: the simulator reacts like an expert checking facts rather than a user holding a position. Prior work found such gaps in simulated answers and queries \citep{aher2023,ma2026,balog2026}; we see a related pattern in reactions, though we did not measure what simulators or users know, and the knowledge bound in \S5 did not reduce disputes of challenges judged wrong, which a pure knowledge account would predict. For sycophancy benchmarks, a GPT-simulated user applies pressure mainly when the model is wrong, which tests fact-checking rather than holding a position under social disagreement. Since Sol~6.1 and the Claude simulators come closer to the real stance effect, sensitivity should be checked per simulator. Future work should separate the knowledge bound from the stance instruction, and randomise stance and correctness for real participants and simulators.

\section*{Limitations}
\paragraph{Observational, selected data.} Stance is not randomised, and C and E probes come from different selection pipelines and differ in depth, period and topic. Within the largest single-pipeline batch, $D$ is smaller but still negative (Appendix~A). C probes are clear, explicit challenges, so $\Dreal$ applies to clear challenges. Stance labels are LLM-assigned and partly noisy (Appendix~A); such label noise biases $\Dreal$ toward zero rather than away from it.
\paragraph{Only users who replied.} Real users leave more often after a challenge, while simulators rarely decline; the all-probes $D$ is estimated on 300 probes only.
\paragraph{LLM judging LLM text.} The judge is the same model as the Sol~6.1 simulator and from the same family as Luna; we collected no human labels under our own codebook. Of two other judges, one reproduces the sign of $D$ and one gives $D$ near zero (Appendix~A). The judge is conservative, and it misses some terse real re-sends of a challenged request (Appendix~A), which makes the pushback-only $D$ conservative, while the repair boundary works the other way. All text checks in \S4 were coded by one LLM reader (Claude), not by a human, with no inter-rater check.
\paragraph{Pushback definition.} Doubting questions (``are you sure?'') count as pushback; recoding them leaves $D$ unchanged (Appendix~A).
\paragraph{One data source.} All real conversations come from WildChat; replicating on LMSYS-Chat-1M \citep{zheng2024}, as \citet{liu2025} do for user feedback, is future work.
\paragraph{Access and versions.} Simulators and judge were reached through the Codex CLI and a third-party API relay; versions cannot be pinned and temperature may not apply. Sol~6.1 was run in two batches, which give different paired estimates (Appendix~A).
\paragraph{Simulators.} Claude simulators were run on 200 conversations only, with one prompt family. The correctness split relies on LLM judgments ($\kappa=0.52$ between two judges) and was not planned in advance; the main correctness judge (gpt-6.1-sol) is from the same family as Luna, so ``judged incorrect'' partly means incorrect by that model family's own knowledge. The Opus-labelled column in Table~\ref{tab:2} is our check on this. Simulators may have seen WildChat in training, but no simulated turn reproduces the real next turn (Appendix~A).

\section*{Ethical Considerations}
WildChat is released under ODC-BY with personal information removed. We do not release conversation text; we will release conversation identifiers, labels, simulated turns, prompts and code. Figure~\ref{fig:1} paraphrases one conversation; no conversation text is quoted.

\section*{Acknowledgements}
LLMs were used to label stance and reactions (\S3), to simulate users, to code the text checks in \S4 (Claude, acting as an auditor; no human coding), and to assist with drafting and editing this paper. The author checked all text and numbers.

\bibliography{refs}

\appendix
\section{Robustness and Judge Details}
\textbf{Table A1 (robustness, Luna).} Regression-adjusted $-0.13$ [$-0.18$, $-0.08$]; user-clustered $-0.12$ [$-0.17$, $-0.08$]; one conversation per user (1,180 users) $-0.15$; inclusion-weighted $-0.12$ [$-0.18$, $-0.06$]; reply-length terciles $-0.05$ (n.s.) / $-0.16$ / $-0.13$; ``are you sure?'' recoded $-0.12$; by assistant model gpt-3.5 $-0.15$, gpt-4o $-0.12$, other GPT-4 $-0.04$ [$-0.14$, $+0.06$]; split-half noise SD 0.046. By single year (exploratory): 2023 $-0.14$ [$-0.21$, $-0.07$]; 2024 $-0.09$ [$-0.16$, $-0.02$]; 2025 (January--July) $-0.25$ [$-0.37$, $-0.13$], n = 86 C / 139 E. Under pushback $\cup$ repair the correctness split remains (Luna 40.8\% / 73.4\% after challenges judged correct / incorrect, 19.2\% / 62.0\% after endorsements; real 24.5\% / 31.9\% and 9.4\% / 15.9\%).

\textbf{Table 2 with CIs.} Real: 19.3 [16.2, 22.5], 23.8 [17.8, 29.7], 7.2 [5.3, 9.3], 12.2 [7.9, 16.4]; Luna: 21.1 [18.5, 23.7], 48.8 [42.3, 55.1], 16.2 [13.8, 18.6], 58.1 [52.3, 64.0].

\textbf{Checks cited in Limitations.} Largest single-pipeline batch (572 C / 217 E): $D=-0.08$ [$-0.16$, $-0.00$]. Relabelling 380 candidate probes moved 16\% to no stance. A Claude Opus~5.5 judge gives the same sign of $D$ on 100 conversations; a Claude Sonnet~5.5 judge without extended thinking agrees less with ours on real turns ($\kappa=0.53$) and gives $D=+0.02$ ($n=100$). Our judge missed 4 of 20 short real turns that re-send a challenged request (read by the LLM auditor). The maximum character 5-gram Jaccard between a simulated and the real next turn is 0.44 (1,383 turns). Sol~6.1 was simulated in two batches: the first 600 conversations (230 earlier ones, all from 2023 H1, plus 370 random) give paired $D_{\text{Sol}}-D_{\text{Luna}}=+0.02$ [$-0.04$, $+0.07$], the other 1,189 give $+0.12$ [$+0.08$, $+0.17$]; by half-year period the estimate ranges from $+0.03$ (2024 H2) to $+0.24$ (2025 H1). We cannot separate sampling variation from drift between the two runs.

\textbf{Judge details.} Single-turn vs.\ mixed judging of real turns: $\kappa=0.66$. Public multiturn-feedback set \citep{liu2025}: 420 turns, coarse $\kappa=0.41$. IARD \citep{cai2020}: precision $\approx0.86$, recall 0.45.

The stance and reaction codebooks, prompts, sample flow and judge checks will be released with our code and labels.

\end{document}